# On the motion of satellite around the natural moons of planets using the concept of ER3BP with variable eccentricity

**Sergey Ershkov\***, <u>Affiliation</u>[1]: Plekhanov Russian University of Economics,

Scopus number 60030998, Russia, e-mail: sergej-ershkov@yandex.ru

<u>Affiliation</u>[2]: Sternberg Astronomical Institute, M.V. Lomonosov's Moscow State University, 13 Universitetskij prospect, Moscow 119992, Russia,

**Dmytro Leshchenko**[3], Odessa State Academy of Civil Engineering and Architecture, Odessa, Ukraine, e-mail: leshchenko_d@ukr.net,

**E.Yu. Prosviryakov**[4], Institute of Engineering Science UB RAS, Ekaterinburg, e-mail: evgen_pros@mail.ru

## Abstract

   In the current study, we explore stability of motion of satellite around the natural moons of planets in Solar system using the novel concept of ER3BP with variable eccentricity. This concept was introduced earlier when novel type of ER3BP (Sun-planet-satellite) was investigated with *variable* spin state of secondary planet correlated implicitly to the motion of satellite (in the synodic co-rotating Cartesian coordinate system) for its trapped orbit near the secondary planet (which is involved in kepler duet "Sun-planet"). But it is of real interest to explore another kind of aforedescribed problem, ER3BP (planet-moon-satellite) with respect to investigation of motion of satellite $m$ around the natural moon $m_{moon}$ of planet in Solar system with variable eccentricity of the moon in its motion around the planet. So, we consider here two primaries, $M_{planet}$ and $m_{moon}$, the last is orbiting around their common barycenter on *quasi-elliptic* orbit with slow-changing, not constant eccentricity (on a large-time scale) due to tidal phenomena. Our aim is to investigate motion of small dot satellite around the natural moon of planet on quasi-stable elliptic orbit. Both novel theoretical and numerical findings (for various cases of trio "planet-moon-satellite") are presented in the current research.

# 1. **Introduction.**

Equations of restricted three-body problem R3BP present the dynamical model of motion of a dot planetoid with small nass under the governing combined action of Newtonian gravity by two large bodies (called primaries in celestial mechanics), dancing in their mutual celestial motion on kepler planar orbits around barycenter. Participants of Celestial mechanics community know a lot of modern and old works presenting outstanding results in R3BP (e.g. [1-7], but not limited to).

In this work, we will assume that small planetoid *m* is orbiting near the natural moon *m moon* of planet in Solar system with variable eccentricity of the moon in its motion around the planet (this kind of motion was considered previously in [8]). So, we consider here two primaries, *M planet* and *m moon*, the last is orbiting around their common barycenter on *quasi-elliptic* orbit with slow-changing eccentricity (on a large-time scale) due to tidal phenomena. Our aim is to investigate motion of small dot satellite around the natural moon of planet on quasi-stable elliptic orbit.

The relative distance $\rho$ of the positions of primaries are always changing in mutual permanent motion (in the ER3BP) on their elliptical orbits [1]

$$\rho = \frac{a_p \cdot (1 - e^2)}{1 + e \cdot \cos f}$$

where, $a_p$ is the semimajor axis of *elliptic* orbits of primaries around their common centre of mass (here, scale of distances is chosen so that $\{a_p(0) * (1 - e(0))\} = 1$), *e* is the *variable* eccentricity, *f* is the *true anomaly*. As formulated in [8], angular motion is given by

$$\frac{df}{dt} = \left( \frac{GM}{a_p^3 \cdot (1 - e^2)^3} \right)^{\frac{1}{2}} \cdot (1 + e \cdot \cos f)^2 \qquad (1)$$

where, *G* is the Gaussian constant of gravitation law, *M* is the sum of masses of primaries, the unit of time is chosen so that constant *G* equals to 1.

According [8-10], equations of ER3BP for small planetoid $m$ can be presented in the synodic co-rotating Cartesian coordinate system $\{x,\ y,\ z\}$ in non-dimensional *scaled* form (we consider here Cauchy problem *with given initial conditions*):

$$\ddot{x}\ -\ 2\,\dot{y}\ =\ \frac{\partial\,\Omega}{\partial\,x}\ ,$$

$$\ddot{y}\ +\ 2\,\dot{x}\ =\ \frac{\partial\,\Omega}{\partial\,y}\ ,\qquad\qquad(2)$$

$$\ddot{z}\qquad\quad =\ \frac{\partial\,\Omega}{\partial\,z}\ ,$$

$$\Omega\ =\ \frac{1}{1+e\cdot\cos f}\left[\frac{1}{2}\Big(x^2+y^2-z^2\cdot e\cdot\cos f\Big)+\frac{(1-\mu)}{r_1}+\frac{\mu}{r_2}\right],\quad(3)$$

where dot indicates derivative with respect to $f$ (here, independent variable) in (2), $\Omega$ is given by expression in (3) the scalar function, and

$$r_1^{\ 2}\ =\ (x-\mu)^2\ +\ y^2\ +\ z^2\ ,$$

$$\qquad\qquad\qquad\qquad\qquad\qquad(4)$$

$$r_2^{\ 2}\ =\ (x-\mu+1)^2\ +\ y^2\ +\ z^2\ ,$$

where $r_i$ $(i = 1, 2)$ are the distances of small planetoid $m$ from primaries $M_{planet}$ and $m_{moon}$, accordingly.

Now, the unit of mass is chosen in (2) so that the sum of the primary masses is equal to 1. We suppose that $M_{planet} \cong 1 - \mu$ and $m_{moon} = \mu$, where $\mu$ is the ratio of the mass of the smaller primary to the total mass of the primaries and $0 < \mu \le 1/2$.

We neglect by the effect of variable masses of the primaries along with by the effect of their *oblateness* [11] as well as by effect of differential rotation on their surfaces [12] (furthermore, we do not take into account the existence of the stable resonance phenomena beween additional moons of host planet [13]). Besides, let us additionally note that the spatial ER3BP when $e > 0$ and $\mu > 0$ is not conservative, and no integrals of motion are known [7].

## 2. Form of analytical presentation of Eqns (2)-(3) convinient for further numercial solving.

According to results of [8], let us present Eqns. (2)-(4) after transforming of their right parts with help of partial derivatives with respect to the proper coordinates as below

$$\ddot{x} - 2\dot{y} = \frac{1}{1 + e \cdot \cos f} \cdot \left[ x - \frac{(1-\mu)(x-\mu)}{\left((x-\mu)^2 + y^2 + z^2\right)^{\frac{3}{2}}} - \frac{\mu(x-\mu+1)}{\left((x-\mu+1)^2 + y^2 + z^2\right)^{\frac{3}{2}}} \right],$$

$$\ddot{y} + 2\dot{x} = \frac{1}{1 + e \cdot \cos f} \cdot \left[ y - \frac{(1-\mu)y}{\left((x-\mu)^2 + y^2 + z^2\right)^{\frac{3}{2}}} - \frac{\mu y}{\left((x-\mu+1)^2 + y^2 + z^2\right)^{\frac{3}{2}}} \right], \quad (5)$$

$$\ddot{z} = \frac{1}{1 + e \cdot \cos f} \cdot \left[ -z \cdot e \cdot \cos f - \frac{(1-\mu)z}{\left((x-\mu)^2 + y^2 + z^2\right)^{\frac{3}{2}}} - \frac{\mu z}{\left((x-\mu+1)^2 + y^2 + z^2\right)^{\frac{3}{2}}} \right],$$

in a form convinient for further numercial solving (in next Sections). We assume that coordinate $z$ of solution $\vec{r} = \{x,\, y,\, z\}$ is associated with the class of *trapped motions*, $z << 1$ (so, planetoid $m$ is to be oscillating close to the plane $\{x, y\}$, $z \cong 0$).

Thus, we can neglect by all the expressions of order $z^2$ and less in (5); so, third equation of system (5) can be transformed as follows

$$\ddot{z} + \omega \cdot z = 0, \quad (6)$$

$$\omega = \frac{1}{1 + e \cdot \cos f} \cdot \left[ e \cdot \cos f + \frac{(1-\mu)}{\left((x-\mu)^2 + y^2\right)^{\frac{3}{2}}} + \frac{\mu}{\left((x-\mu+1)^2 + y^2\right)^{\frac{3}{2}}} \right] \quad (7)$$

Besides, we should remark that equation (6) describes the dynamics of component $z$ of the solution depending on the already known coordinates $\{x, y\}$ and on true anomaly $f$; this is *Riccati*-type [10] ordinary differential equation (with variable coefficient $\omega(f)$, determined in (7)).

Thus far, we conclude that solving procedure for the system of equations (5) is reduced to solving of its two first equations

$$\ddot{x} \; - \; 2\,\dot{y} \; = \; \frac{1}{1 + e \cdot \cos f} \cdot \left[ x - \frac{(1-\mu)(x-\mu)}{\left((x-\mu)^2 + y^2\right)^{\frac{3}{2}}} - \frac{\mu(x-\mu+1)}{\left((x-\mu+1)^2 + y^2\right)^{\frac{3}{2}}} \right],$$

$$\tag{8}$$

$$\ddot{y} \; + \; 2\,\dot{x} \; = \; \frac{y}{1 + e \cdot \cos f} \cdot \left[ 1 - \frac{(1-\mu)}{\left((x-\mu)^2 + y^2\right)^{\frac{3}{2}}} - \frac{\mu}{\left((x-\mu+1)^2 + y^2\right)^{\frac{3}{2}}} \right].$$

But we should especially note additionally that assumption of coordinate $z$ to be approx. zero in (6) corresponds to the *partial* case of the well known Clohessy-Wiltshire equations for relative motion when $e \neq 0$. Numerical solution for coordinate $z$ can be obtained for the already known coordinates $\{x, y\}$, obtained numerically as solutions of Eqns. (8) at previous step of our solving procedure.

### 3. The way of taking into account the variable eccentricity $e(f)$ in Eqns. (5).

We should especially remark (see [8]) that for using of the time-dependent eccentricity $e(t)$ in Eqns (2)-(3) or equivalently in Eqns. (5) we should first solve Eqn. (1) with aim of expressing time $t$ *via* true anomaly $f$ (independent variable in (2)-(3)). Meanwhile, introducing the dependence of eccentricity on true anomaly in Eqns. of ER3BP allows to take into account the effect of tidal phenomena on orbital motions of primaries which are participating in dynamical effects in these equations, during a long-time period. Thus insofar, equations of motion (ER3BP) (5) depend on the aforementioned variable eccentricity.

Equation (1) can be transformed to other form (see [8]) in case of low-eccentricity orbit $e \cong 0$ by neglecting the terms of second order of smallness in (1) as follows

$$f \cong 2\arctan\left(\tan\left(\frac{Ct}{2}\right)(1+2e)\right) \qquad (9)$$

where in (9), $C$ is the constant of integration having dimension inverse to time; besides, eccentricity $e$ is assumed to be very slowly varying function on long-time scale period; so, it could be considered equal to constant *close to zero* for the suffitiently large period of changing of true anomaly $f$. Using expression given in [8], we obtain (here below, $e_0 = e(0)$, $a_0 = a_p(0) = (1-e_0)^{-1}$)

$$e \cong e_0 \cdot \exp\left(-\frac{B}{2} \cdot \frac{\exp\left(\frac{13}{2}e_0^2\right)}{(a_0)^{\frac{13}{2}}} \cdot t\right), \quad \Rightarrow \quad \left\{ B = \frac{21 k_2^m M_{planet}\left(\pm\sqrt{G(M_{planet}+m_{moon})}\right) \cdot \left(R_{moon}\right)^5}{Q^m m_{moon}} \right\}$$

$$\Rightarrow \quad \tan\left(\frac{f}{2}\right) \cong \left(1 + 2e_0 \cdot \exp\left(-\frac{B}{2} \cdot \frac{\exp\left(\frac{13}{2}e_0^2\right)}{(a_0)^{\frac{13}{2}}} \cdot t\right)\right) \cdot \tan\left(\frac{Ct}{2}\right) \qquad (10)$$

where let us remind the denotations in expression for $B$ mentioned above in equation (10): $m_{moon}$ is the mass of moon, $M_{planet}$ is the mass of planet, $\dfrac{k_2^m}{Q^m}$ is the ratio of the Love number $k_2^m$ of moon to its quality factor $Q^m$ (which describes the response of the potential of the distorted body in regard to the influence of current tides); $R_{moon}$ is the equatorial radius of the moon.

Let us also clarify additionally that formulae (9)-(10) were obtained in [8] based on results of work [14] where both host star and planet (here, planet and moon) were assumed to be "rigid-type celestial bodies" for which main contribution influencing on orbit of moon in its motion around the host planet stems from the tides raised on the moon by the planet (see conclusion in [13] in regard to this matter) that alter the exchange of angular momentum between the bodies.

Whereas in "fluid-type planet" (such as Jupiter) we should use another modification of formula (10) as follows (here below, $e_1 = e_0$, $a_1 = a_0$ just for simplicity for choosing initial data in both scenario; besides, $(A/C) << 1$)

$$e \cong e_1 \cdot \left( \frac{39}{2} \cdot A \cdot \left( a_1 \right)^{-\frac{13}{2}} \cdot t \right)^{\frac{19}{52}} \Rightarrow \left\{ A = \frac{k_2 \, m_{moon} \left( \pm \sqrt{G \left( M_{planet} + m_{moon} \right)} \right) \cdot R^5}{Q \, M_{planet}} \right\} \Rightarrow$$

$$\tan \left( \frac{f}{2} \right) \cong \left( 1 + 2e_1 \cdot \left( \frac{39}{2} \cdot A \cdot t \right)^{\frac{19}{52}} \right) \cdot \tan \left( \frac{Ct}{2} \right) \Rightarrow t \cong \frac{f}{C} \Rightarrow e \cong e_1 \cdot \left( 1 + \left( \frac{39 \cdot 19}{104} \right) \cdot \frac{A}{C} \cdot f \right) \quad (11)$$

where main contribution influencing on orbit of moon in its motion around the host planet stems from the tides raised on the surface of planet by the moon orbiting in the 1:1 spin–orbit resonance around planet insofar their mutual Newton attraction is still actual (here, $R$ is equatorial radius of the planet; $\frac{k_2}{Q}$ is the ratio of the Love number $k_2$ of planet to its quality factor $Q$). Mathematical procedure of derivation of Eq. (11) has also been moved to an Appendix, with only the resulting formulae left in the main text.

Thus, we obtain from (10) the approximate equation which can be solved by applying of series of Taylor expansions (by neglecting the terms of second order of smallness like $e_0^2$ or $(B \cdot e_0)$ since $B \to 0$ due to $R_{moon} \to 0$ in expression for $B$ if scaling with respect to $a_p (0)$, e.g. for Moon: $R_{moon} \cong 0.0045$) as follows

$$\tan\left(\frac{f}{2}\right) \cong \left(1 + 2e_0\left(1 - \frac{B}{2}(1 - \frac{13}{2}e_0)t\right)\right) \cdot \tan\left(\frac{Ct}{2}\right) \;\; \Rightarrow \;\; \tan\left(\frac{f}{2}\right) \cong \left(1 + 2e_0\right) \cdot \tan\left(\frac{Ct}{2}\right)$$

$$\Rightarrow \quad \begin{cases} t \cong \dfrac{f}{\left(1 + 2e_0\right) \cdot C} & \left(\{Ct, f\} \to \pi \cdot |k|, \;\; k \in Z\right) \\[4mm] t \cong \dfrac{f}{C} & \left(\{Ct, f\} \in \left(\dfrac{\pi \cdot n}{4}, \dfrac{\pi \cdot n}{2}\right), \;\; n \in N\right) \end{cases} \qquad (12)$$

Thus far, we can make a conclusion from (10) and (12) regarding *form of dependence* of eccentricity of moon in Eqns. (5) on true anomaly (at least, up to the terms of second order of smallness with respect to *f*):

$$e \cong e_0 \cdot \exp\left(-\frac{B}{2C} \cdot \frac{\exp\left(\frac{13}{2}e_0^2\right)}{(a_0)^{\frac{13}{2}}\left(1 + 2e_0\right)} \cdot f\right), \;\; \Rightarrow \;\; e \cong e_0 \cdot \left(1 - \frac{B}{2C} \cdot f\right) \qquad (13)$$

(we should choose sign "minus" for expression for *B* in (10), where (*B/C*) << 1).

So, we can use the semi-analytical expression for function *e*(*f*) (13) for obtaining numerical solution of Eqns. (5) or of system of Eqns. (6-8).

## 4. <u>Discussion & Conclusion.</u>

As we can see from the aforepresented results and solving procedure, the introducing of the dependence of eccentricity on true anomaly in equations of ER3BP allows definitely taking into consideration the dynamical effect of tidal phenomena (of long-time period) on orbital evolution of primaries in these equations. It is worth noting the essential restriction to our model such that we have consuderd the case of *small eccentricity*.

In the current study, we explore stability of motion of satellite around the natural moons of planets in Solar system using the novel concept of ER3BP with variable eccentricity. This concept was introduced earlier when novel type of ER3BP (Sun-planet-satellite) was investigated with *variable* spin state of secondary planet correlated implicitly to the motion of satellite (in the synodic co-rotating Cartesian coordinate system) for its trapped orbit near the secondary planet (which is involved in kepler duet "Sun-planet"). But it is of real interest to explore another kind of aforedescribed problem, ER3BP (planet-moon-satellite) with respect to investigation of motion of satellite $m$ around the natural moon $m_{moon}$ of planet in Solar system with variable eccentricity of the moon in its motion around the planet. So, we consider here two primaries, $M_{planet}$ and $m_{moon}$, the last is orbiting around their common barycenter on *quasi-elliptic* orbit with slow-changing, not constant eccentricity (on a large-time scale) due to tidal phenomena. Our aim is to investigate motion of small dot satellite around the natural moon of planet on quasi-stable elliptic orbit. Both novel theoretical and numerical findings (for various cases of trio "planet-moon-satellite") are presented in the current research.

Let us provide the numerical calculations of the approximated solutions for system (8), where we consider $e(f) = e_0 \cdot (1 + \Omega \cdot f)$, $\Omega = 0.0001 \, [rad^{-1}]$, and $z^2 \to 0$.

We should note that we have used for numerical code the Runge–Kutta 4th-order method with step 0.001 starting from initial conditions. We have chosen for our numercial calculations (for modelling the triple system "Earth – Moon – satellite" $\{M_{planet}, m_{moon}, m\}$) as follows:

$$e_0 = 0.0549, \ \mu \cong (7.36*10^{22}/5.9742*10^{24}) \sim 1.232*10^{-2}.$$

As for the initial data, we have chosen as follows: 1) $x_0 = -1.01$, $(x')_0 = 0$; 2) $y_0 = 0$, $(y')_0 = 0$. Results of numerical calculations are imagined on Figs.1-4.

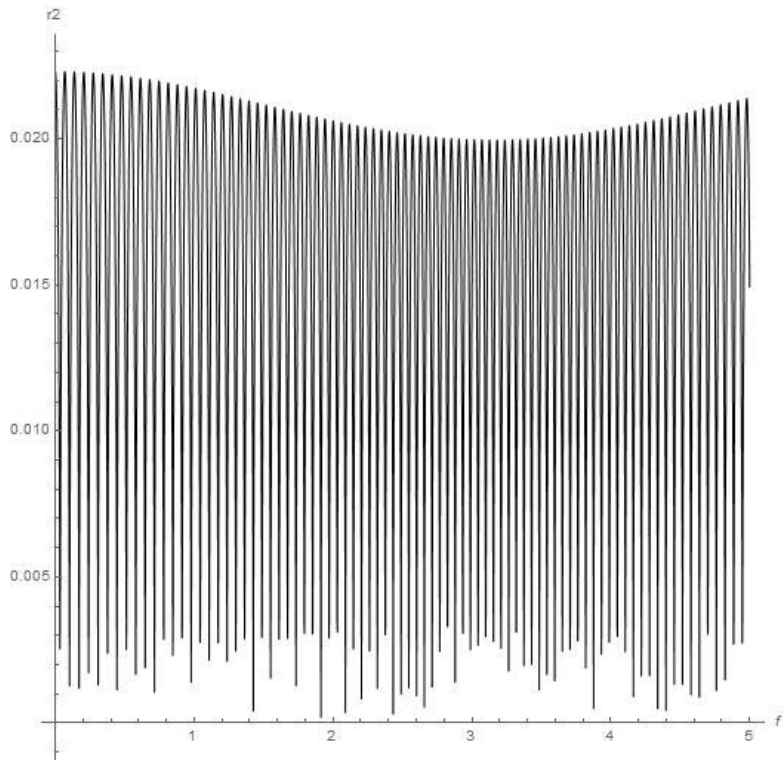

Fig.1. Results of numerical calculations of the $r_2$.

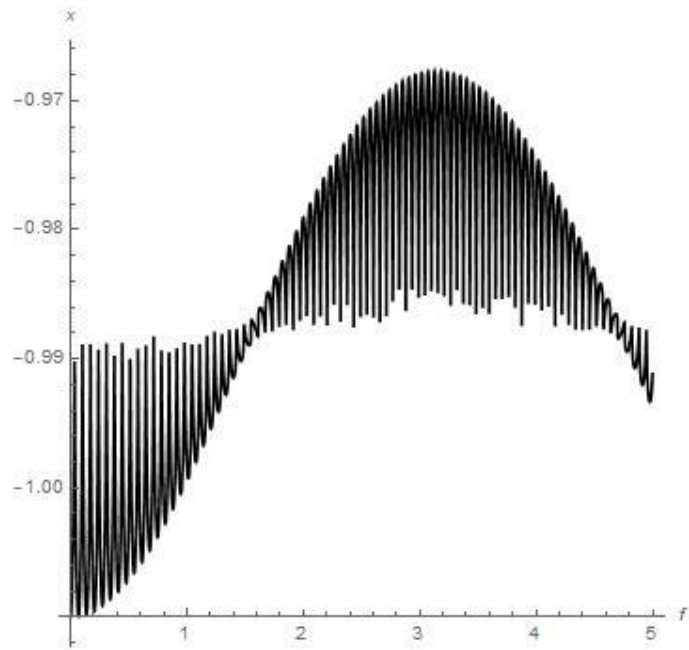

Fig.2. Results of numerical calculations of the coordinate $x$.

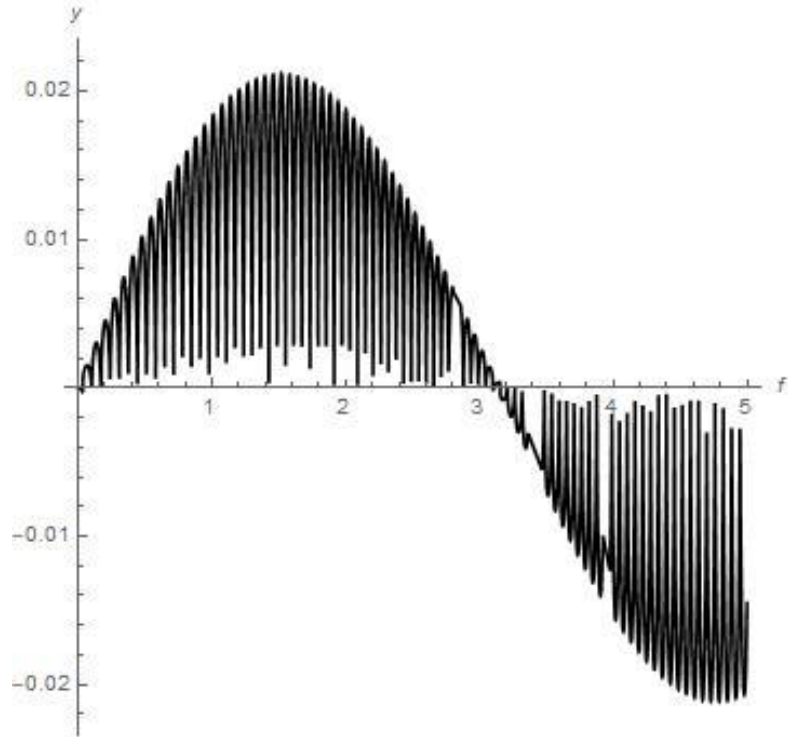

Fig.3. Results of numerical calculations of the coordinate *y*.

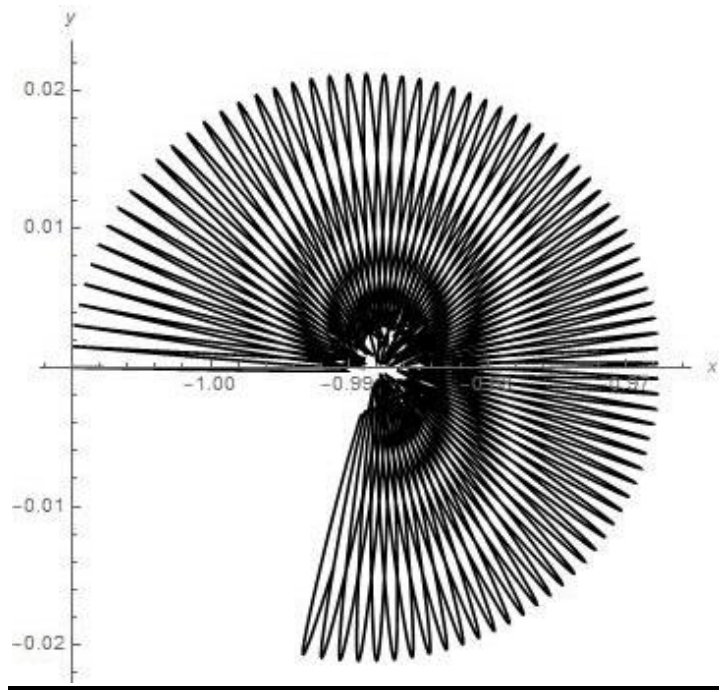

Fig.4. Results of numerical calculations for trajectory in {*x*, *y*} plane.

Meanwhile, the numerical approximation for the dynamics of the infinitesimal planetoid $m$ in this case (see Figs.1-4) means that this small celestial body experiences strong oscillations which may disrupt such small satellite in a mechanical way due to exceeding the body tensile strength a much when moving near the Moon in system "Earth-Moon-satellite" ($r_2 << 1$) in the aforepointed catastrophic oscillating regime. Also we should remark that the dynamical behaviour for the components of the solution is unstable. We have tested various sets of initial data, but there are no stable results for numerical experiments. This reliable attempts confirm that the dynamics for the components of the solution is unstable, indeed. This should mean that small satellite $m$ does not have stable orbit even at value of true anomaly $f \cong 5$ (Fig.1) (or less than 1 full turn of Moon around the Earth starting from initial point). It is worthnoting to demonstrate stable character of coordinate $z(f)$ on Fig.5 (in a sense of oscillations near the $\{x, y, 0\}$ plane), numerically obtained for solutions plotted on Figs. 1-4; we have chosen the initial data as follows: $z_0 = 0.001$, $(z')_0 = 0$ (the range of true anomaly is chosen up to the meaning $f = 30$ which can be surely continued further):

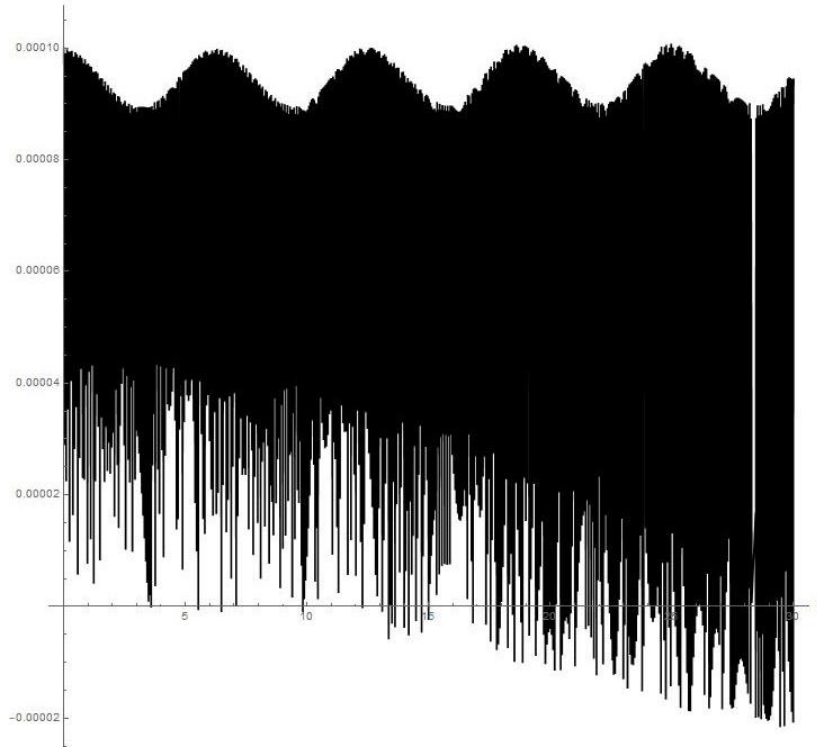

Fig.5. Plot for coordinate $z(f)$ using Eqns. (6-7) ($z$ on ordinate, $f$ on abscissa axis).

Let us discuss Figs. 1-4 further by disclosing more detailed analysis regarding these graphical plots of numerical solutions:

-   Fig.1 presents the results of numerical calculations for the distance $r_2$ of planetoid $m$ from the Moon. Namely, we can see from Fig.1 that distance $r_2$ is strongly oscillating (there are 15 peaks of oscillations during first 1 radian of changing $f$) in a narrow range $f \in$ [0.0002; 0.0225] with obvious further approx. stable regime during a long period of changing the true anomaly $f$, up to the value $f = 5$ (circa 1 full turns of Moon around the Earth starting from initial point) and further;

-   Figs.2-3 present the results of numerical experiment for the coordinates $\{x, y\}$. We can see that coordinates $\{x, y\}$ are strong oscillating each in a narrow range of true anomaly $f$ (e.g., coordinate $x$ experiences 16 peaks of oscillations during first 1 radian of changing $f$ which corresponds to less than 1/6 of full angular turn of Moon around the Earth starting from initial point);

-   Fig.4 presents numerical calculations for trajectory of planetoid in $\{x, y\}$ plane. We can see that small satellite is apparently strong oscillating (e.g., the trajectory has 51 peaks of oscillations in a top half-plane);

It is also worthnoting to continue the dynamics presented on Fig. 1 up to meaning $f = 100$ on Fig.6, where can also see the strong oscillations up to the meaning of true anomaly $f = 100$ (which corresponds circa to the 16 full turns of orbit's of the Moon around the Earth starting from initial point).

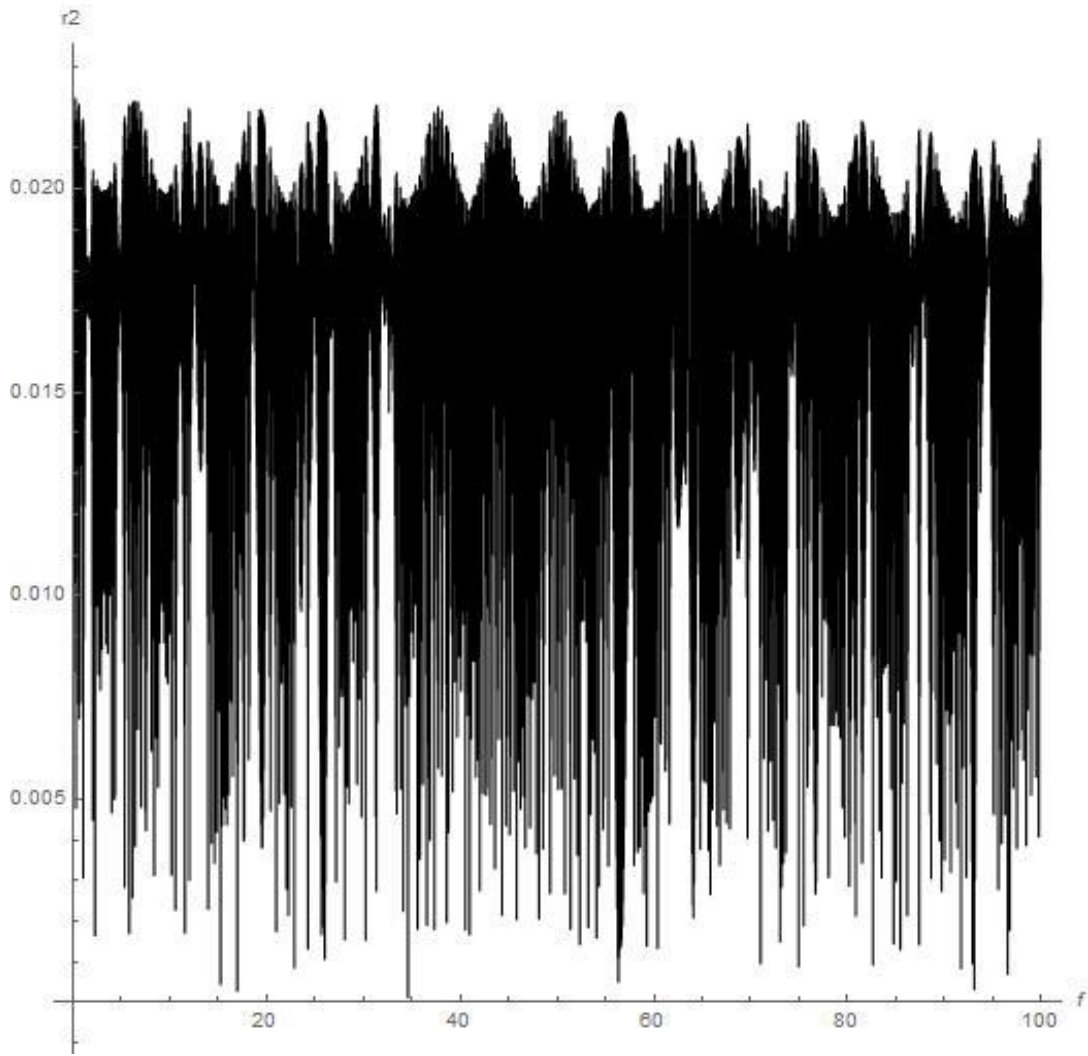

Fig.6. Results of numerical calculations of the $r_2$ (up to meaning $f = 100$).

The last but not least, we should especially outline that we have investigated cases of moons of Mars (Fig.7-8) and one of four large satellites of Jupiter, Callisto on Fig.9 (there are 4 large satellites of Jupiter: Ganymede, Callisto, Io and Europa, but 3 of them: Ganymede, Io and Europa are known to be captured in the *Laplace resonance* [13] and so, beyond of the pure formulations and aims of our research). As we can see, there are no stable dynamics for satellite of moons for all these cases, except trio "Jupiter-Callisto-planetoid".

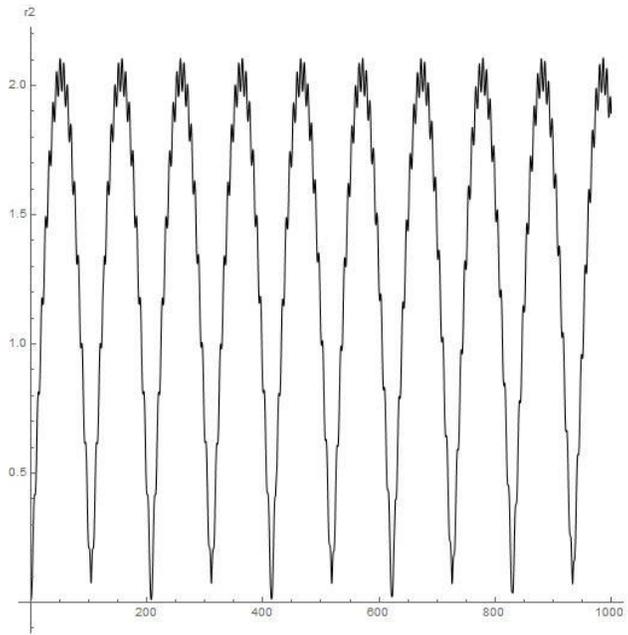

Fig.7. Results of numerical calculations of the $r_2$ for Phobos, moon of Mars (up to meaning $f = 1000$). Initial conditions are chosen as previously on Figs.1-3;

$e_0 = 0.0151$, $\mu \cong (1.072*10^{16}/6.42*10^{24}) \sim 0.167*10^{-8}$.

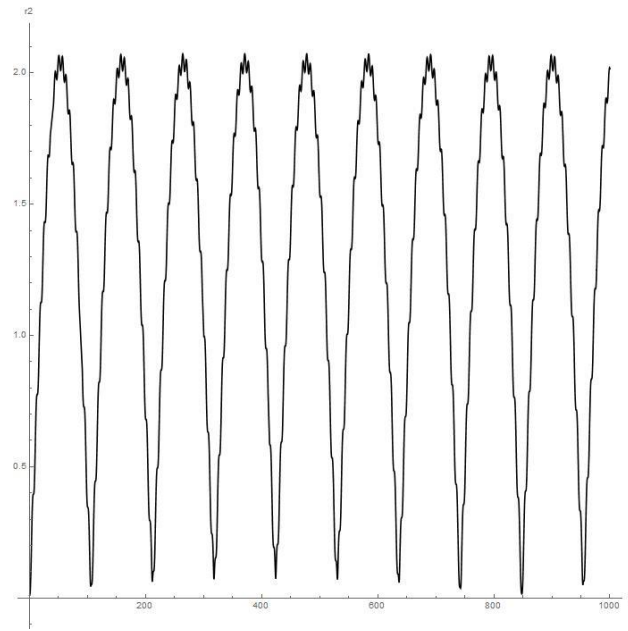

Fig.8. Results of numerical calculations of the $r_2$ for Deimos, moon of Mars (up to meaning $f = 1000$). Initial conditions are chosen as previously on Figs.1-3;

$e_0 = 0.0002$, $\mu \cong (1.48*10^{15}/6.42*10^{24}) \sim 0.231*10^{-9}$.

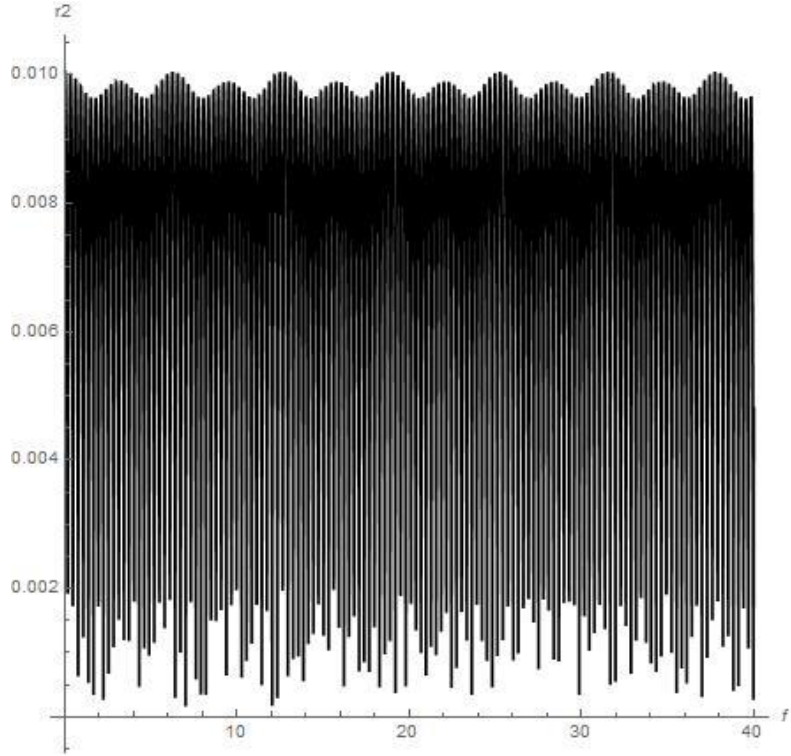

Fig.9. Results of numerical calculations of the $r_2$ for Callisto, moon of Jupiter (up to meaning $f = 1000$). Initial conditions are chosen as previously on Figs.1-3; $e_0 = 0.0074$, $\mu \cong (1.075*10^{23}/1.9*10^{27}) \sim 0.5658*10^{-4}$.

As we can see from Figs. 7-8, dynamics of dot satellite near each of moons of Mars is unstable, whereas the numerical approximation for the dynamics of the infinitesimal planetoid $m$ in case of Jupiter (Fig.9) means that this small celestial body experiences strong oscillations which may disrupt such small satellite in a mechanical way due to exceeding the body tensile strength a much when moving near the Callisto in system "Jupiter-Callisto-satellite" ($r_2 << 1$) in the aforementioned catastrophic oscillating regime (maximal close approach of planetoid to Callisto in {$x, y, 0$} plane is circa $1.5*10^{-4}$). It is worthpointing out to demonstrate stable character of coordinate $z(f)$ on Fig.10 (in a sense of oscillations near the {$x, y, 0$} plane), numerically obtained for solutions plotted on Fig. 9 with respect to planetoid moving near the Callisto in system "Jupiter-Callisto-satellite"; we have chosen the initial data as previously: $z_0 = 0.001$, $(z')_0 = 0$ (the range of true anomaly is chosen up to the meaning $f = 30$ which can be surely continued further):

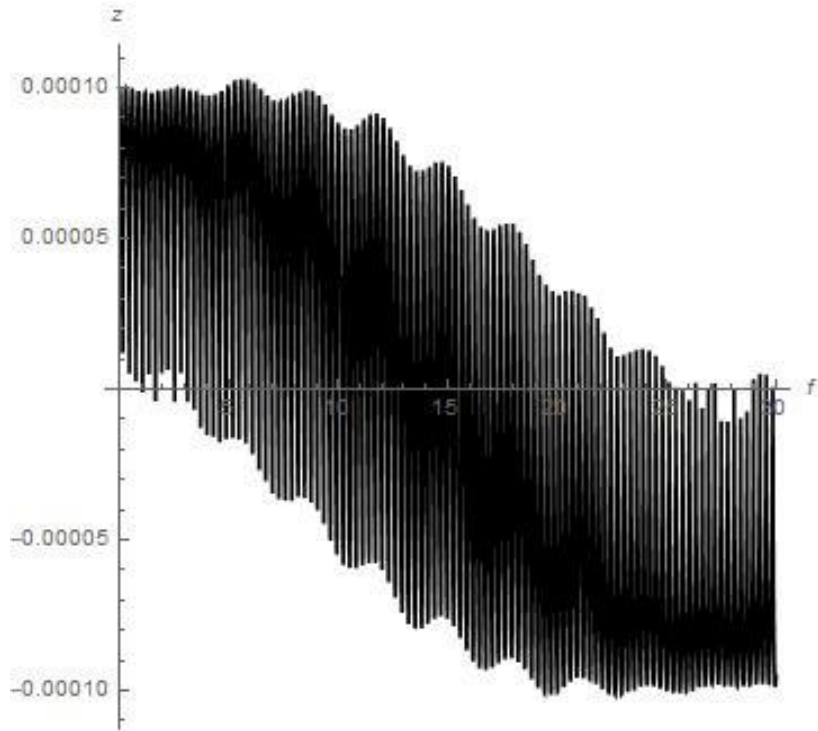

Fig.10. Plot for coordinate $z(f)$ using Eqns. (6-7) ($z$ on ordinate, $f$ on abscissa axis) for planetoid moving near the Callisto in system "Jupiter-Callisto-satellite".

Also, the remarkable additional articles should be cited, which concern the problem under consideration, [14-33]. To finalize conclusions, let us remark that there is an old competetive discussion between members of celestial mechanics community which concept should be used:

1) constant Q for tidal evolution (not our choice; for example, see [34]);
2) approximation assuming "constant time lag" in the equilibrium tide model of tidal friction (this model was used e.g. in REBOUND integrator, see [35]);
3) the quality factor Q of the Planet is assumed to be depending on the tidal-flexure frequency (our choice definitely; see [14]).

Futhermore, modern approaches to orbital motions with tides (e.g., see REBOUND integrator [35]) use numerical schemes that evaluate all the perturbations/forces at each timestep and can evolve a system "Planet-moon-satellite" for millions of years ($\sim 10^9$ submoon orbits). Moreover, these numerical approaches also consider the tides raised on the host bodies (e.g., planet and moon) by the host star that alter the exchange of angular momentum between the bodies. The current work includes these extra forces for possible consideration in case of "Jupiter-Callisto-satellite" (since Jupiter should be considered as "fluid-type planet" with another type of expression (11) presenting the dependence of low eccentricity $e$ on true anomaly $f$).

So, in this research we do present the actual and novel algorithm for calculating the submoon orbit in conception of ER3BP for trio "Planet-moon-satellite" (with slowly-changing low eccentricity of the moon) both for "rigid-type planet" such as Earth, Mars (10) and "fluid-type planet" Jupiter (11) confirming the utility/accuracy of the derived expressions for such orbit.

## **Conflict of interest**

On behalf of all authors, the corresponding author states that there is no conflict of interest.

Remark regarding contributions of authors as below:

In this research, Dr. Sergey Ershkov is responsible for the general ansatz and the solving procedure, simple algebra manipulations, calculations, results of the article and also is responsible for the search of analytical and semi-analytical solutions.
Prof. Dmytro Leshchenko is responsible for theoretical investigations as well as for the deep survey in literature on the problem under consideration.

Dr. Evgeniy Prosviryakov is responsible for obtaining numerical solutions related to approximated ones (including their graphical plots).

All authors agreed with results and conclusions of each other in Sections 1-4.

## **Appendix (mathematical procedure of derivation of Eq. (11)).**

Let us present mathematical procedure of derivation of Eq. (11) as follows:

Since we consider that main contribution influencing on orbit of moon in its motion around the host planet stems from the tides raised *on the surface of planet* by the moon orbiting in the 1:1 spin–orbit resonance around the planet, we can use formulae (3.2) obtained in [14] e.g. dynamical invariant which interrelates semimajor axis with respect to the eccentricity ($\{a_1, e_1\} = \{a_p(0), e(0)\} = const$):

$$a_p = a_1 \cdot \left(\frac{e}{e_1}\right)^{\frac{8}{19}} \cdot \exp\left(\frac{51}{19}(e^2 - e_1^2)\right), \qquad (A.1)$$

where the term: $\exp((51/19)\cdot(e^2 - e_1^2)) \cong 1$. For the reason that time *t* has not been involved to be presented in expressions in both parts of ($A.1$), we can change the independent variable $t \to f$ in ($A.1$) for $\{a_p(t),\ e(t)\} \to \{a_p(f), e(f)\}$ (and *vice versa*) without losing a generality for the dynamical invariant ($A.1$). Let us present Eqn. (1) in other form as below

$$\frac{df}{dt} = \left(\frac{GM}{a_1^3 \cdot \left(\frac{e}{e_1}\right)^{\frac{24}{19}} \cdot \exp\left(\frac{153}{19}(e(f)^2 - e_1^2)\right) \cdot (1 - e(f)^2)^3}\right)^{\frac{1}{2}} \cdot (1 + e(f) \cdot \cos f)^2 \qquad (A.2)$$

which can be transformed in case of low-eccentricity orbit $e \cong 0$ (by neglecting of terms of second order smallness in $(A.2)$) as follows

$$\frac{d\,f}{d\,t} = \left(\frac{e}{e_1}\right)^{-\frac{12}{19}} \cdot \left(\frac{GM}{a_1^{\,3}}\right)^{\frac{1}{2}} \cdot (1 + e(f) \cdot \cos f)^2 \quad \Rightarrow$$

$$\left\{ C = \left(\frac{GM}{a_1^{\,3}}\right)^{\frac{1}{2}}, \quad \left(\frac{e}{e_1}\right)^{-\frac{12}{19}} \cong 1 \right\} \quad \Rightarrow \quad \int \frac{d\,f}{(1 + 2e \cdot \cos f)} \cong C t \qquad (A.3)$$

(let us remind that we have chosen $e_1 = e_0$, $a_1 = a_0$ in (11) just for simplicity of presentation the final result). Where in in $(A.3)$ eccentricity $e$ is very slowly varying function on long-time scale period; so, it could be considered equal to constant in Eqn $(A.3)$ for the suffitiently large period of changing of true anomaly $f$. Thus, we have obtained in $(A.3)$ the equation solution of which approximately results to (9) as follows (see [8]):

$$\int \frac{d\,f}{(1 + 2e \cdot \cos f)} \cong C t \quad \Rightarrow \quad \frac{2}{\sqrt{1 - 4e^2}} \arctan\left(\frac{(1 - 2e)\tan(f/2)}{\sqrt{1 - 4e^2}}\right) \cong C t \quad \Rightarrow$$

$$2\arctan\left((1 - 2e)\tan(f/2)\right) \cong C t \quad or \quad f \cong 2\arctan\left(\tan\left(\frac{Ct}{2}\right)(1 + 2e)\right)$$